\documentclass[conference]{IEEEtran}
\IEEEoverridecommandlockouts
\usepackage{cite}
\usepackage{amsmath,amssymb,amsfonts}
\usepackage{graphicx}
\usepackage{textcomp}
\usepackage{xcolor}

\usepackage{amsfonts,amssymb}
\usepackage{amsmath, bm}
\usepackage{algorithm}
\usepackage{algorithmicx}
\usepackage{algpseudocode}
\usepackage{makecell}
\usepackage{stfloats}
\usepackage{subfigure}
\usepackage{mathrsfs}

\def\BibTeX{{\rm B\kern-.05em{\sc i\kern-.025em b}\kern-.08em
    T\kern-.1667em\lower.7ex\hbox{E}\kern-.125emX}}
\begin{document}

\newcommand{\B}{\mathbf}
\newcommand{\T}{^\mathsf{T}}
\newcommand{\CT}{^\mathsf{H}}

\title{RIS-Assisted Secure Transmission Exploiting Statistical CSI of Eavesdropper}

\author{\IEEEauthorblockN{Cen Liu, Chang Tian, and Peixi Liu}
\IEEEauthorblockA{State Key Laboratory of Advanced Optical Communication Systems and Networks, \\
Department of Electronics, Peking University, Beijing, China \\
Email: \{cen, tianch, liupeixi\}@pku.edu.cn}
}

\maketitle

\begin{abstract}
We investigate the reconfigurable intelligent surface (RIS) assisted downlink secure transmission where only the statistical channel of eavesdropper is available. To handle the stochastic ergodic secrecy rate (ESR) maximization problem, a deterministic lower bound of ESR (LESR) is derived. We aim to maximize the LESR by jointly designing the transmit beamforming at the access point (AP) and reflect beamforming by the phase shifts at the RIS. To solve the non-convex LESR maximization problem, we develop a novel penalty dual convex approximation (PDCA) algorithm based on the penalty dual decomposition (PDD) optimization framework, where the exacting constraints are penalized and dualized into the objective function as augmented Lagrangian components. The proposed PDCA algorithm performs double-loop iterations, i.e., the inner loop resorts to the block successive convex approximation (BSCA) to update the optimization variables; while the outer loop adjusts the Lagrange multipliers and penalty parameter of the augmented Lagrangian cost function. The convergence to a \textit{Karush-Kuhn-Tucker} (KKT) solution is theoretically guaranteed with low computational complexity. Simulation results show that the proposed PDCA scheme is better than the commonly adopted alternating optimization (AO) scheme with the knowledge of statistical channel of eavesdropper.
\end{abstract}


\section{Introduction}
\IEEEPARstart{R}{econfigurable} intelligent surface (RIS), which has given rise to the brand-new concept of ``smart radio environment", is regarded as a promising technology in the future 5G beyond and 6G cellular network\cite{TiejunCui14, MarcoDiRenzo20, QurratUlAinNadeem20, QingqingWuTWC, HuayanGuo20, ShuhaoZeng21}. RIS is comprised of many uniformly distributed low-cost reflecting elements that can adjust both amplitude and phase of the impinging signal, so as to form a directional beam in a passive manner. It is compatible with existing wireless communication systems and can bring many distinctive advantages, such as low power consumption, noise-free and self-interference-free under full-duplex mode. Therefore, RIS-assisted wireless transmission systems are getting growing interest from both academia and industry.

Physical layer security, as a key technology aiming at providing confidential message transmission and solving the privacy protection problem in the physical layer, has received considerable attention in the past decade\cite{AmitavMukherjee14, AshishKhisti10}. For the enhancement of security performance, a basic approach is to improve the spectral efficiency of the legitimate users or deteriorate that of the eavesdroppers (Eve). In an RIS-assisted wireless system, the received signal can be suppressed at the Eve while being boosted at the user. Thus, the deployment of RIS can bring a new degree of freedom (DoF) in the space domain, and the security performance can be further improved.

Several works related to the RIS-assisted physical layer security have emerged recently \cite{MiaoCui19,HongShen19,XianghaoYu19,DongfangXu19,XinrongGuan20,WeihengJiang20}. In \cite{MiaoCui19,HongShen19} and \cite{XianghaoYu19}, an RIS-assisted single-user single-Eve multi-input single-output (MISO) communication system was investigated with or without taking the direct links of access point (AP)-user and AP-Eve into account, respectively. Artificial noise was jointly designed with transmit beamforming and RIS phase shifts both in the multi-user single-Eve \cite{DongfangXu19} and single-user multi-Eve MISO systems \cite{XinrongGuan20}. Furthermore, the security performance of an RIS-assisted multi-input multi-output (MIMO) system was analyzed in \cite{WeihengJiang20}. Existing literature usually assume the global instantaneous channel state information (CSI) is perfectly known, which is impractical in many cases. At first, the Eve-related instantaneous CSI is hard to be fully available since it is impractical to implement the Eve-related channel estimation in a real-time manner due to the uncertainty of Eve's position. In addition, the CSI between RIS and mobile user can not be estimated directly by traditional channel estimation techniques due to the RIS's passive nature.

This paper is dedicated to the investigation of RIS-assisted secure transmission where only the statistical CSI of Eve is available. The main contributions are summarized as follows:
\begin{enumerate}
	\item The scenario we consider is more practical than the works mentioned above since the statistical CSI is relatively easy to obtain through long-term observation;
	\item A lower bound of the ergodic secrecy rate is derived to cope with the formulated stochastic optimization problem by resorting to its more tractable deterministic counterpart;
	\item The proposed penalty dual convex approximation (PDCA) algorithm can provably solve the induced non-convex rate maximization problem to a \textit{Karush-Kuhn-Tucker} (KKT) point. Meanwhile, it possesses lower computational complexity and better performance than the commonly adopted alternating optimization (AO) algorithm.
\end{enumerate}

\emph{Notations:} $\left(\cdot\right)\CT$, $\left(\cdot\right)\T$, and $\left(\cdot\right)^*$ stand for the Hermitian, the transpose and the conjugate operations, respectively. $|\cdot|$ and $\|\cdot\|$ represent the modulus of a complex number and the $\mathcal{\ell}_2$ norm of a complex vector. ${\rm diag}\left(\cdot\right)$ denotes the diagonal matrix whose diagonals are the elements of the input vector.

\begin{figure}[t]
	\centerline{\includegraphics[width=0.8\columnwidth]{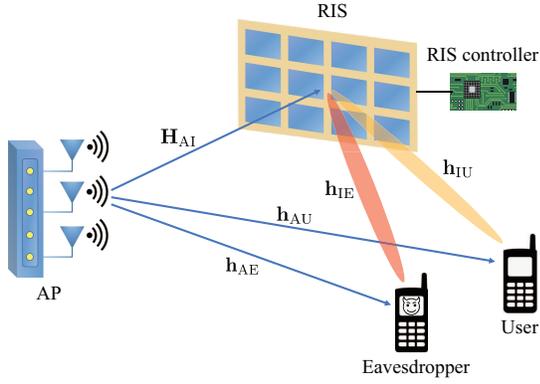}}
	\caption{RIS-assisted single-user MISO wireless system in the presence of a single-antenna eavesdropper.}
	\label{fig1}
\end{figure}

\section{System Model}

We consider an RIS-assisted single-user MISO system in the presence of a single-antenna eavesdropper, as illustrated in Fig. \ref{fig1}. The AP and RIS are equipped with $M$ antennas and $N$ reflecting elements, respectively. The AP transmits confidential message to the legitimate single-antenna user through the direct link $\B{h}_{\rm AU}\in\mathbb{C}^M$ and the reflecting link with the help of RIS. The reflecting link can be viewed as a cascaded channel of AP-RIS and RIS-user channel which are denoted by $\B{H}_{\rm AI}\in\mathbb{C}^{N\times M}$ and $\B{h}_{\rm IU}\in\mathbb{C}^N$, respectively. The received signal at the user is
\begin{align}
	y_{\rm U} &= \left(\B{h}_{\rm IU}\CT \B\Phi \B{H}_{\rm AI} + \B{h}_{\rm AU}\CT\right)\B{w} s + n_{\rm U}
\end{align}
where $\B\Phi = {\rm diag}\left(\boldsymbol\phi\right)$ , $\boldsymbol\phi = \left[e^{j\theta_1}, \ldots, e^{j\theta_N}\right]\T$ denotes the phase shifts induced by each reflecting element of RIS. $\B{w} \in \mathbb{C}^M$ denotes the transmit beamforming vector at the AP. $s$ and $n_{\rm U}$ denote the transmitted information and additive white Gaussian noise (AWGN) at the user, respectively, with $\mathbb{E}\left[|s|^2\right]=1$ and $n_{\rm U}\sim\mathcal{CN}\left(0,\sigma_U^2\right)$. It can be easily verified that $\B{h}_{\rm IU}\CT \B\Phi \B{H}_{\rm AI} = \boldsymbol\phi\T \B{H}_{\rm U}$ where $\B{H}_{\rm U} = {\rm diag}\left(\B{h}_{\rm IU}^*\right)\B{H}_{\rm AI}$ stands for the cascaded channel from the AP to RIS to the user. Due to the passive nature of RIS, the RIS-user channel $\B h_{\rm IU}$ can hardly be obtained. Nevertheless, the cascaded channel $\B H_{\rm U}$ can be estimated more easily\cite{ZhenqingHe20, WenhuiZhang21}. We assume the user-related channels $\B{h}_{\rm AU}$, $\B{H}_{\rm U}$ and $\B{H}_{\rm AI}$ are perfectly known at the AP and experience quasi-static flat fading.

Similar to the user, the signal received by Eve is
\begin{align}
	y_{\rm E} &= \left(\B{h}_{\rm IE}\CT \B\Phi \B{H}_{\rm AI} + \B{h}_{\rm AE}\CT\right)\B{w} s + n_{\rm E}
\end{align}
where $\B h_{\rm IE}$ is the RIS-Eve channel and $\B h_{\rm AE}$ is the AP-Eve channel. $n_{\rm E}\sim\mathcal{CN}\left(0,\sigma_{\rm E}^2\right)$ is AWGN at the Eve. We can also replace $\B{h}_{\rm IE}\CT \B\Phi \B{H}_{\rm AI}$ in (2) by $\boldsymbol\phi\T \B{H}_{\rm E}$, where $\B{H}_{\rm E} = {\rm diag}\left(\B{h}_{\rm IE}^*\right)\B{H}_{\rm AI}$ stands for the cascaded channel from the AP to RIS to the Eve. We denote the AP-Eve channel $\B h_{\rm AE}$ and the RIS-Eve channel $\B h_{\rm IE}$ as follows as in\cite{YuHan19}
\begin{align}
	\B h_{i{\rm E}} &= \sqrt{\zeta_0\left(\frac{d_{i{\rm E}}}{d_0}\right)^{-\alpha_{i{\rm E}}}} \left(\sqrt{\frac{K_{i{\rm E}}}{K_{i{\rm E}}+1}} \bar{\B h}_{i{\rm E}} + \sqrt{\frac{1}{K_{i{\rm E}}+1}} \tilde{\B h}_{i{\rm E}}\right), \notag \\
	i &\in \{\rm A,\rm I \}.
\end{align}
where $\zeta_0$ denotes the path loss at reference distance $d_0$, $d_{i{\rm E}}$ is the distance between AP/RIS and Eve. $\alpha_{i{\rm E}}$ and $K_{i{\rm E}}$ denote path loss exponent and Rician $K$-factor of corresponding channels. $\bar{\B h}_{i{\rm E}}$ represents the line-of-sight (LoS) component which is determined by the angle of departure (AoD) at the AP/RIS and angle of arrival (AoA) at the Eve, while the non-line-of-sight (NLoS) component is denoted by $\tilde{\B h}_{i{\rm E}} \sim \mathcal{CN} (0,\B I)$. The statistical CSI of Eve that can be taken advantage of are the means and covariance matrices of the AP-Eve channel $\B h_{\rm AE}$ and the RIS-Eve channel $\B h_{\rm IE}$ since they are relatively easy to obtain\cite{JunZhang20}.

\section{Problem Formulation}

\subsection{Ergodic Secrecy Rate Maximization}
The secrecy rate from AP to user has the following expression
\begin{align}
	R_{\rm S} = \left(R_{\rm U} - R_{\rm E}\right)^+
\end{align}
where $\left(\cdot\right)^+ \triangleq \max\{\cdot,0\}$ is the non-negative operator, and
\begin{align}
	R_{\rm U} = \log_2 \left(1 + \frac{\left|\left(\boldsymbol\phi\T \B{H}_{\rm U} + \B{h}_{\rm AU}\CT \right) \B{w}\right|^2}{\sigma_{\rm U}^2}\right)\\
	R_{\rm E} = \log_2 \left(1 + \frac{\left|\left(\boldsymbol\phi\T \B{H}_{\rm E} + \B{h}_{\rm AE}\CT \right) \B{w}\right|^2}{\sigma_{\rm E}^2}\right)
\end{align}
are the achievable rates at the user and Eve, respectively. Taking expectation over Eve's statistical channels, we get the ergodic secrecy rate (ESR) as follows
\begin{align}
	\bar{R}_{\rm S} = \mathbb{E}_{\mathcal{H}_{\rm E}} \left[ \left(R_{\rm U} - R_{\rm E}\right)^+ \right]
\end{align}
where $\mathcal{H}_{\rm E} \triangleq \{\B h_{\rm AE}, \B h_{\rm IE}\}$. Then the ESR maximization problem can be formulated as
\begin{subequations}
	\begin{align}
		\max_{\boldsymbol\phi, \B w} \ & \bar{R}_{\rm S}\\
		{\rm s.t.} \ & |\phi_i| = 1,\ i=1,2,\ldots,N,\\
		&\|\B w\|^2 \leq P_{\rm max}.
	\end{align}
\end{subequations}
where (8b) and (8c) are unit-modulus constraint of RIS and maximum transmit power constraint at the AP, respectively.

\subsection{Problem Reformulation}
With the expectation operator in the objective (8a), this stochastic optimization problem can not be efficiently handled. Therefore, we convert it into a more tractable form by firstly deriving a deterministic lower bound of the ergodic secrecy rate (LESR), which is given by the following proposition.

\emph{Proposition 1:} The ESR of the RIS-assisted MISO system $\bar{R}_{\rm S}$ is lower bounded by $\bar{R}_{\rm S}^{\rm lb}$ given in (9) which is placed on the top of this page, where $\B G_{\rm A}$, $\B G_{\rm I}$ and $\B G_{\rm AI}$ are some parameters given by
\setcounter{equation}{9}
\begin{align}
	\B G_i &= \zeta_0 \left(\frac{d_{i{\rm E}}}{d_0}\right)^{-\alpha_{i{\rm E}}}\left(\frac{K_{i{\rm E}}}{K_{i{\rm E}}+1} \bar{\B h}_{i{\rm E}} \bar{\B h}_{i{\rm E}}\CT + \frac{1}{K_{i{\rm E}}+1}\B I\right), \notag \\
	i &\in \{\rm A,\rm I \}.
\end{align} 
\begin{align}
	\B G_{\rm AI} = &\zeta_0 \left( \frac{d_{\rm AE}}{d_0} \right)^{-\frac{\alpha_{\rm AE}}{2}} \left( \frac{d_{\rm IE}}{d_0} \right)^{-\frac{\alpha_{\rm IE}}{2}} \notag \\
	&\times \sqrt{\frac{K_{\rm AE} K_{\rm IE}}{(K_{\rm AE} + 1)(K_{\rm IE}+1)}} \bar{\B h}_{\rm AE} \bar{\B h}_{\rm IE}\CT
\end{align}

\emph{Proof:} See Appendix A.$\hfill\blacksquare$ 

\begin{figure*}[t]
	\begin{align} 
		\bar{R}_{\rm S}^{\rm lb} =
		\left[\log_2 \left(\frac{\frac{1}{\sigma_{\rm U}^2} \left|\left(\boldsymbol\phi\T \B H_{\rm U} + \B h_{\rm AU}\CT \right) \B w\right|^2 + 1}{\frac{1}{\sigma_{\rm E}^2} \left(\|\B G_{\rm A}^\frac{1}{2} \B w\|^2 + \|\B G_{\rm I}^\frac{1}{2} \B\Phi \B H_{\rm AI} \B w\|^2 + 2\Re\left[\B w\CT \B G_{\rm AI} \B\Phi \B H_{\rm AI} \B w\right]\right) + 1} \right)\right]^+ \tag{9}
	\end{align}
	\hrulefill
\end{figure*}

By \emph{Proposition 1}, the stochastic ESR maximization problem can be addressed by solving the following deterministic LESR maximization problem
\begin{align}
	\max_{\B w, \boldsymbol\phi} \ & \bar{R}_{\rm S}^{\rm lb}\\
	{\rm s.t.}\ &\text{(8b),\ (8c).} \notag
\end{align}
Nonetheless, problem (12) is very challenging to solve since it is highly non-convex caused by the coupling optimization variables in the objective function and the non-convex constraint (8b). By utilizing the novel penalty dual decomposition (PDD) framework\cite{QingjiangShi20, YunlongCai18}, we propose an efficient PDCA algorithm which can provably solve problem (12) to KKT points.

\section{The Proposed Penalty Dual Convex Approximation Algorithm}
Intuitively, AO can be applied to solve problem (12) as in \cite{MiaoCui19,HongShen19,XianghaoYu19,DongfangXu19,XinrongGuan20,WeihengJiang20}. But to ensure the convergence of the block coordinate descent (BCD) method (AO is a special case of BCD since there are only a pair of optimization variables), the subproblem of each block variable need to be constrained in a convex set and also to be solved to its unique global optimal\cite{MeisamRazaviyayn13}. As to primal problem (12), the former requirement can not be satisfied due to the non-convex constraint (8b) while the latter is hard to meet since the objective of each subproblem is non-convex, so the global optimality of the solutions can not be guaranteed. Specifically, semidefinite relaxation (SDR) as adopted in\cite{MiaoCui19,DongfangXu19} and\cite{XinrongGuan20} can only provide an approximate local optimal solution.

To address above issues, in this section we propose a penalty dual convex approximation (PDCA) algorithm by employing the penalty dual decomposition (PDD) optimization framework in which the sequence of solution are guaranteed to converge to a KKT point\cite{QingjiangShi20, YunlongCai18}. By applying the PDD technique, we first obtain an
alternative formulation to (12) where an augmented
Lagrangian term is incorporated into the objective in order
to handle the unit-modulus constraint. We then develop an block successive convex approximation (BSCA) algorithm to solve the augmented Lagrangian problem in the inner loop of the PDD framework.  Finally, we summarize the PDCA
joint beamforming algorithm and analyze its
computational complexity.

At the beginning, we drop the non-negative operator in the objective of (12) since zero rate can be achieved at least by adopting $\B w = \B 0$. Also, the outer logarithmic function is omitted because of its monotonicity.

\subsection{Augmented Lagrangian Problem}
To start with, an augmented Lagrangian (AL) function with penalty parameter $\varrho$ is denoted by $\mathcal{L}_\varrho (\boldsymbol\phi, \B w;\boldsymbol\lambda)$ while the dual variable $\boldsymbol\lambda$ is introduced to deal with the constraint (8b). Further, let us define an \textit{AL problem} ($P_{\varrho,\boldsymbol\lambda}$) as follows
\begin{align}
	(P_{\varrho,\boldsymbol\lambda}) \ \max_{\boldsymbol\phi, \B w\in \mathcal{W}} \  \bigg\{&\mathcal{L}_\varrho (\boldsymbol\phi, \B w;\boldsymbol\lambda) \triangleq \bar{R}_{\rm S}^{\rm lb}(\boldsymbol\phi, \B w)\notag\\
	- \frac{1}{2\varrho}&\sum_{i=1}^{N} \left[ (|\phi_i|-1-\varrho \lambda_i)^2 - (\varrho \lambda_i)^2 \right] \bigg\}
\end{align}
where $\mathcal{W} = \left\{ \B w \ |\ \|\B w\|^2 \leq P_{\rm max}\right\}$. The proposed PDCA algorithm exhibits a double-loop
structure, where the outer loop updates the dual variable and
the penalty parameter while the inner loop seeks to optimize
the primal variables by solving problem ($P_{\varrho,\boldsymbol\lambda}$). The key to the
PDD method is the inner iterations for solving augmented
Lagrangian problems.

\subsection{Proposed BSCA Algorithm for Solving AL Problem}
In the following, we develop an BSCA algorithm
to solve ($P_{\varrho,\boldsymbol\lambda}$) in the inner loop of the PDD optimization
framework. With only the convex constraint (8c) left, BCD-type algorithms can be used to solve problem ($P_{\varrho,\boldsymbol\lambda}$) to ensure the accessibility of a KKT solution of the primal problem (12) as suggested in\cite{QingjiangShi20}. Here, an inexact variant of BCD named block successive convex approximation (BSCA)\cite{MeisamRazaviyayn13} is adopted to solve above AL problem. Following a simple cyclic rule, convex approximation of the objective of each block variable is updated successively and in each iteration a sufficient decrease of the objective value can be yielded. By fixing the transmit beamformer in ($P_{\varrho,\boldsymbol\lambda}$), the RIS phase shifts design problem is stated as
\begin{align}
	\min_{\boldsymbol\phi}\ \bigg\{h(\boldsymbol\phi) \triangleq &-\frac{\boldsymbol\phi\CT \B C \boldsymbol\phi + 2\Re\left[ \boldsymbol\phi\CT \B c_1 \right] + c_2}{\boldsymbol\phi\CT \B D \boldsymbol\phi + 2\Re\left[ \boldsymbol\phi\CT \B d_1 \right] + d_2}\notag\\
	&+ \frac{1}{2\varrho}\sum_{i=1}^{N}\left( |\phi_i|-1-\varrho \lambda_i \right)^2\bigg\}
\end{align}
where
\begin{align}
	&\B C = \frac{1}{\sigma_{\rm U}^2} \B H_{\rm U}^* \B w^* \B w\T \B H_{\rm U}\T,\notag\\
	&\B D = \frac{1}{\sigma_{\rm E}^2} {\rm diag}(\B H_{\rm AI}^* \B w^*) \B G_{\rm I} {\rm diag}(\B H_{\rm AI} \B w),\notag\\
	&\B c_1 = \frac{1}{\sigma_{\rm U}^2} \B H_{\rm U}^* \B w^* \B w\T \B h_{\rm AU}^*,\ \B d_1 = \frac{1}{\sigma_{\rm E}^2} {\rm diag}(\B H_{\rm AI}^* \B w^*) \B G_{\rm AI}\CT \B w,\notag\\
	&c_2 = \frac{1}{\sigma_{\rm U}^2} \left|\B h_{\rm AU}\CT \B w\right|^2 + 1,\ d_2 = \frac{1}{\sigma_{\rm E}^2} \|\B G_{\rm A}^\frac{1}{2} \B w\|^2 + 1.
\end{align}
A strongly convex quadratic function $\hat{h}(\boldsymbol\phi, \boldsymbol\phi^k)$ is selected as a surrogate function in the neighborhood of the $k$th iteration point $\boldsymbol\phi^k$
$$\hat{h}(\boldsymbol\phi, \boldsymbol\phi^k) \triangleq h(\boldsymbol\phi^k) + \left< h'(\boldsymbol\phi^k), \boldsymbol\phi - \boldsymbol\phi^k \right> + \frac{1}{2\alpha^{k+1}}\|\boldsymbol\phi - \boldsymbol\phi^k\|^2$$
who satisfies $\hat{h}'(\boldsymbol\phi^k,\boldsymbol\phi^k) = \frac{\partial \mathcal{L}_\varrho (\boldsymbol\phi^k, \B w;\boldsymbol\lambda)}{\partial\boldsymbol\phi}$ (Assumption (6.1) in \cite{MeisamRazaviyayn13}). The next iteration is generated by
\begin{align}
	\hat{\boldsymbol\phi}^{k+1} = \arg\min_{\boldsymbol\phi} \hat{h}(\boldsymbol\phi, \boldsymbol\phi^k) = \boldsymbol\phi^k - \alpha^{k+1}\nabla h (\boldsymbol\phi^k)
\end{align}
which is exactly the gradient descent step. Nevertheless, $\hat{\boldsymbol\phi}^{k+1}$ may not be a point of sufficient decrease, which differs from that in block successive upper-bound minimization (BSUM)\cite{MingyiHong16} since $\hat{h}$ is only a local approximation of $\mathcal{L}_\varrho$ in BSCA. For this reason, backtracking line search approach with the \textit{Armijo} step size selection rule\cite{NumericalOptimization} is exploited to search for a point of sufficient decrease, denoted by $\boldsymbol\phi^{k+1}$.

Fixing the RIS phase shifts in ($P_{\varrho,\boldsymbol\lambda}$) results in following transmit beamforming design problem at the AP
\begin{align}
	\min_{\B w \in \mathcal{W}} \ \bigg\{g(\B w) \triangleq -\frac{\B w\CT \B A \B w + 1}{\B w\CT \B B \B w + 1} \bigg\}
\end{align}
where $\B A = \frac{1}{\sigma_{\rm U}^2} \B a\B a\CT$ with $\B a = (\boldsymbol\phi\T \B H_{\rm U} + \B h_{\rm AU}\CT)\CT$ and $	\B B = \frac{1}{\sigma_{\rm E}^2} (\B G_{\rm A} + \B B_1\CT \B G_{\rm I} \B B_1 + \B B_2  + \B B_2\CT) $ with $\B B_1 = \B\Phi \B H_{\rm AI}$, $\B B_2 = \B G_{\rm AI} \B B_1$. The problem (17) can be addressed by the same approach as stated above, with the only difference is that gradient projection, rather than gradient descent as (16), is performed as follows 
\begin{align}
	\hat{\B w}^{k+1} =\mathbb{P}_\mathcal{W}\left[ \B w^k - \alpha^{k+1}\nabla g (\B w^k) \right]
\end{align}
where $\mathbb{P}_\mathcal{W}\left[ \cdot \right]$ denotes the projection onto the convex set $\mathcal{W}$.

\subsection{Proposed PDCA Algorithm and Computational Complexity}
The proposed PDCA algorithm performs double-loop iterations: the inner loop resorts to the BSCA to update the optimization variables; while the outer loop adjusts the Lagrange multipliers and penalty parameter of the AL function. The PDCA algorithm is summarized in Algorithm \ref{alg1}. In this algorithm, the notation `optimize($P_{\varrho^r, \boldsymbol\lambda^r}, \boldsymbol\phi^r, \B w^r$)' represents the optimization oracle which using the BSCA, as depicted in Algorithm \ref{alg2}, to iteratively solve the AL problem ($P_{\varrho^r, \boldsymbol\lambda^r}$) based on the current iteration point \{$\boldsymbol\phi^r, \B w^r$\}. Compared with the SDR-based AO algorithm proposed in\cite{MiaoCui19} whose worst-case computational complexity is $\mathcal{O}(I_\text{AO}I_\text{SDR}N^{4.5})$\cite{ZhiquanLuo10}, that of our double-loop PDCA algorithm is only $\mathcal{O}(I_\text{PDD}I_\text{BSCA} N^2)$, where $I_{\left[ \cdot \right]}$ denotes the number of iterations of the corresponding algorithm.

\begin{algorithm}[t]
	\caption{PDCA Algorithm for Problem (12)}\label{PDCA}
	\begin{algorithmic}[1]
		\State \textbf{initialization:} Set $r=0$, constraint violation tolerance $\eta>0$, error tolerance $\epsilon>0$, initialize feasible \{$\boldsymbol\phi^0$, $\B w^0$\}, $\boldsymbol\lambda^0$, $\varrho^0>0$ and assign $\bar{R}_{\rm S}^{{\rm lb}(0)} = \infty$.
		\Repeat
		\State $\{\boldsymbol\phi^{r+1}, \B w^{r+1}\} = optimize(P_{\varrho^r, \boldsymbol\lambda^r}, \boldsymbol\phi^r, \B w^r)$
		\State \textbf{if} $\||\boldsymbol\phi^{r+1}|-\B 1\|_\infty \leq \eta$
		\State \qquad $\boldsymbol\lambda^{r+1} = \boldsymbol\lambda^r + \frac{1}{\varrho^r}\left( |\boldsymbol\phi^{r+1}| -\B 1 \right)$
		\State \qquad $\varrho^{r+1} = \varrho^r$
		\State \textbf{else}
		\State \qquad $\boldsymbol\lambda^{r+1} = \boldsymbol\lambda^r$
		\State \qquad update $\varrho^{r+1}$ by decreasing $\varrho^r$
		\State \textbf{end}
		\State $r=r+1$
		\State Calculate $\bar{R}_{\rm S}^{{\rm lb}(r)} = \bar{R}_{\rm S}^{\rm lb}\left(\boldsymbol\phi^r, \B w^r\right)$
		\Until $\left|\bar{R}_{\rm S}^{{\rm lb}(r)} - \bar{R}_{\rm S}^{{\rm lb}(r-1)}\right| \leq \epsilon$. 
	\end{algorithmic}
	\label{alg1}
\end{algorithm}

\begin{algorithm}[t]
	\caption{BSCA Algorithm for the AL Problem $(P_{\varrho,\boldsymbol\lambda})$ }\label{BSCA}
	\begin{algorithmic}[1]
		\State \textbf{input:} $\varrho^r$, $\boldsymbol\lambda^r$, $\boldsymbol\phi^r$, $\B w^r$ given by Algorithm 1.
		\State \textbf{initialization:} Set $k=0$, error tolerance $\epsilon'>0$, \{$\boldsymbol\phi^0, \B w^0$\} = \{$\boldsymbol\phi^r, \B w^r$\}, $\boldsymbol\lambda = \boldsymbol\lambda^r$, $\varrho = \varrho^r$, $\alpha_1^{\text{ini}}, \alpha_2^{\text{ini}}>0$, $\rho_1, \rho_2\in (0,1)$, $c_1, c_2\in (0,1)$.
		\Repeat
		\State $\alpha_1 = \alpha_1^{\text{ini}}$, $\alpha_2 = \alpha_2^{\text{ini}}$
		\Repeat\qquad \textit{// backtracking line search}
		\State $\alpha_1 = \rho_1 \alpha_1$
		\Until $h(\hat{\boldsymbol\phi}^{k+1}) \leq h(\boldsymbol\phi^k) - c_1\alpha_1  \|\nabla h(\boldsymbol\phi^k)\|^2$
		\State $\boldsymbol\phi^{k+1} = \hat{\boldsymbol\phi}^{k+1}$
		\Repeat\qquad \textit{// backtracking line search}
		\State $\alpha_2 = \rho_2 \alpha_2$
		\Until $g(\hat{\B w}^{k+1}) \leq g(\B w^k) - c_2\alpha_2  \|\nabla g(\B w^k)\|^2$
		\State $\B w^{k+1} = \hat{\B w}^{k+1}$
		\State $k=k+1$
		\State Calculate $\mathcal{L}_\varrho^k = \mathcal{L}_\varrho (\boldsymbol\phi^k, \B w^k;\boldsymbol\lambda)$
		\Until $\left| \mathcal{L}_\varrho^k - \mathcal{L}_\varrho^{k-1} \right| \leq \epsilon'$
		\State \textbf{output:} $\{\boldsymbol\phi^{r+1}, \B w^{r+1}\} = \{\boldsymbol\phi^k, \B w^k\}$
	\end{algorithmic}
	\label{alg2}
\end{algorithm}

\section{Simulation Results}

\begin{figure}[t]
	\centerline{\includegraphics[width=0.9\columnwidth]{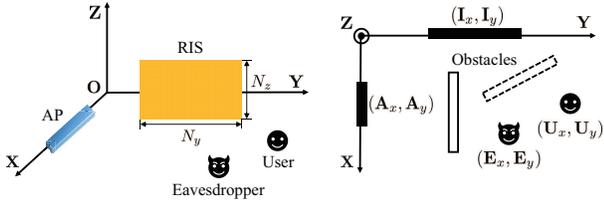}}
	\caption{Simulation setup.}
	\label{fig2}
\end{figure}

\begin{table}[t]
	\centering
	\caption{Simulation Parameters}
	\begin{tabular}{|l|l|}
		\hline
		\textbf{Parameter} & \textbf{Value}\\
		\hline
		\text{Number of antennas} & $M=8$.\\
		\hline
		\makecell[l]{\text{Normalized antenna}\\ \text{or element separation}} & $\Delta_{\rm A} = \Delta_{\rm I} = \frac{1}{2}$.\\
		\hline
		\text{Path loss at 1m} & $\zeta_0 = -30$\text{dB}.\\
		\hline
		\text{Path loss exponent} & \makecell[l]{$\alpha_{\rm AU}=\alpha_{\rm AE}=3.67$,\\ $\alpha_{\rm IU}=\alpha_{\rm IE}=2.2$, $\alpha_{\rm AI}=2$.}\\
		\hline
		\text{Rician} $K$\text{-factor} & \makecell[l]{$K_{\rm AU}=K_{\rm AE}=0$,\\ $K_{\rm IU}=K_{\rm IE}=10^{0.9}$, $K_{\rm AI}=\infty$.}\\
		\hline
		\text{Noise power} & $\sigma_{\rm U}^2=\sigma_{\rm E}^2=-90$\text{dBm}.\\
		\hline
		\text{Coordinate/m} & \makecell[l]{$\left({\rm A}_x,{\rm A}_y\right) = \left(5,0\right)$, $\left({\rm I}_x,{\rm I}_y\right) = \left(0,50\right)$, \\ $\left({\rm U}_x,{\rm U}_y\right) = \left(5,60\right)$, $\left({\rm E}_x,{\rm E}_y\right) = \left(10,55\right)$.}\\
		\hline
	\end{tabular}
	\label{tab1}
\end{table}

\begin{figure}[t]
	\centering
	\subfigure[$N=160$]{
		\begin{minipage}[t]{0.465\linewidth}
			\centering
			\includegraphics[width=1.535in]{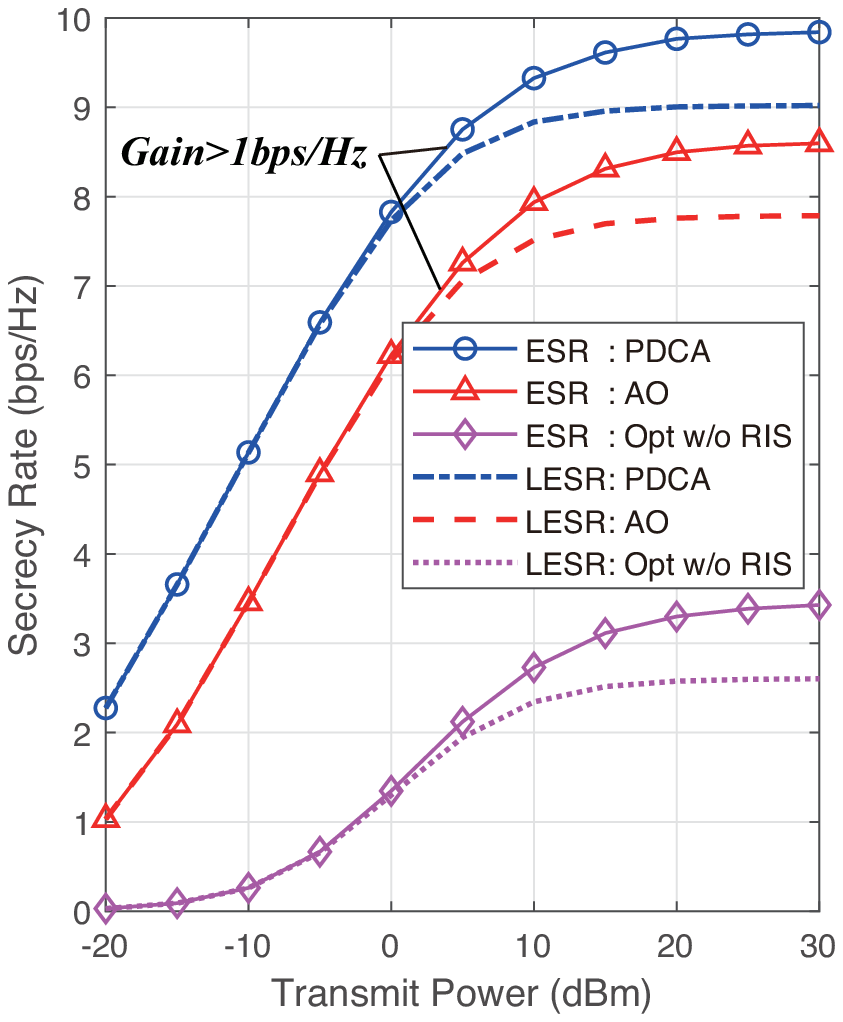}
		\end{minipage}
	}
	\subfigure[$P_{\rm max} = 5$dBm]{
		\begin{minipage}[t]{0.465\linewidth}
			\centering
			\includegraphics[width=1.5in]{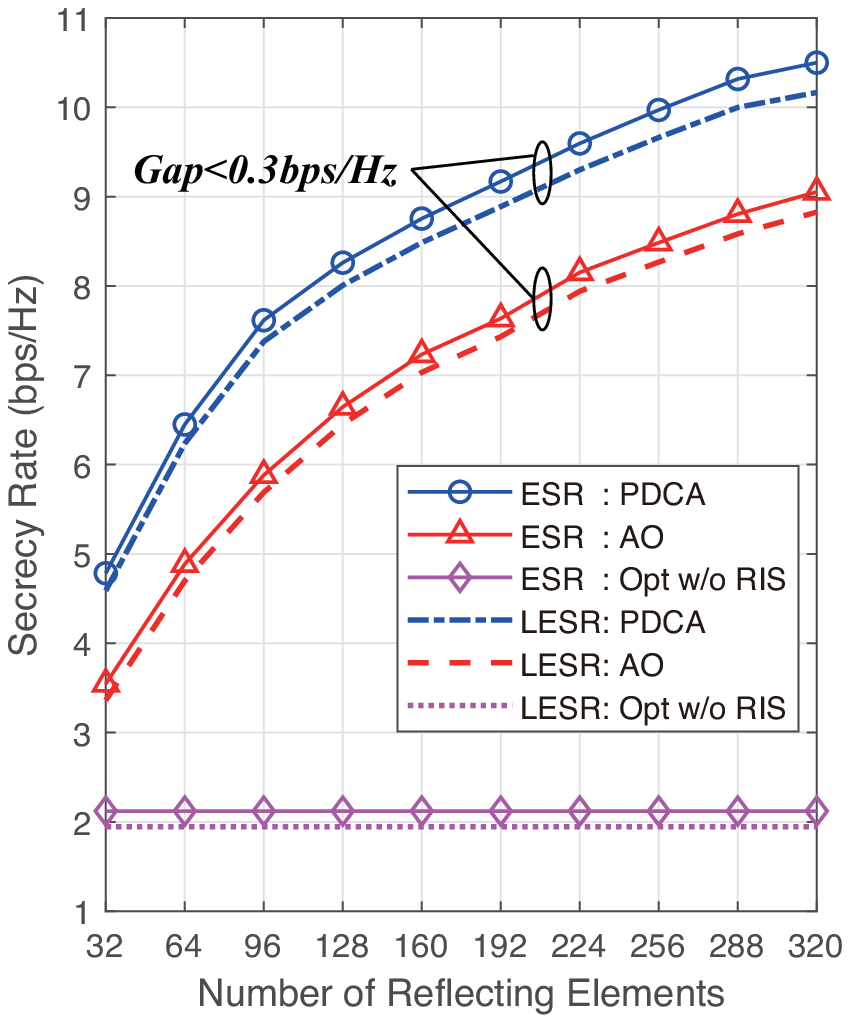}
		\end{minipage}
	}
	\centering
	\caption{ESR and LESR versus $P_{\rm max}$ and $N$.}
	\label{fig3}
\end{figure}

\begin{figure}[t]
	\centering
	\subfigure{
		\begin{minipage}[t]{1\linewidth}
			\centering
			\includegraphics[width=2.7in]{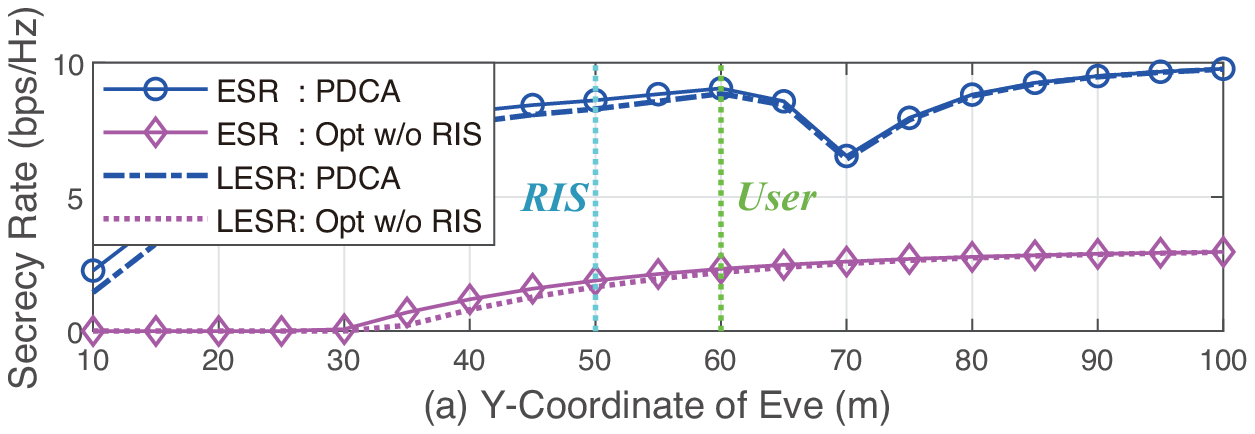}
		\end{minipage}
	}
	\subfigure{
		\begin{minipage}[t]{1\linewidth}
			\centering
			\includegraphics[width=2.7in]{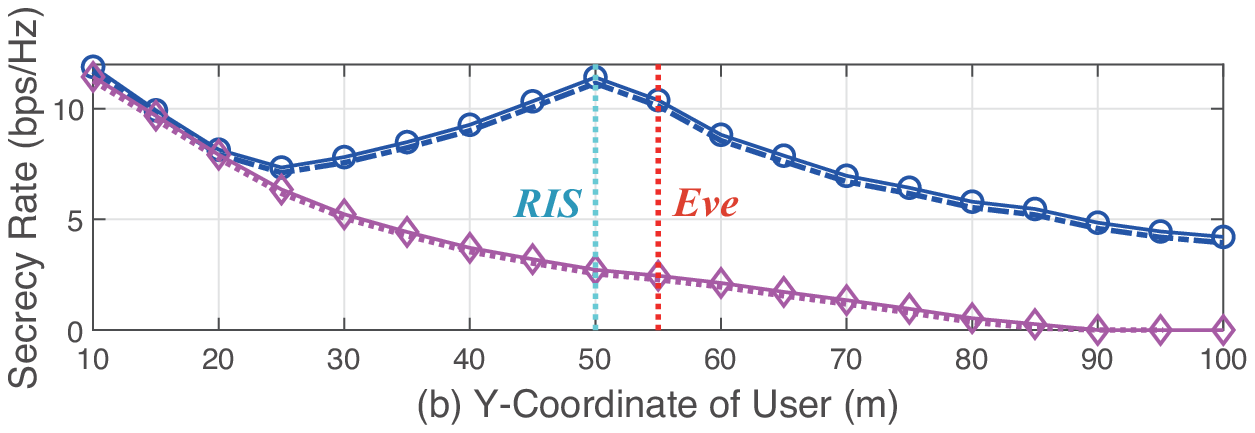}
		\end{minipage}
	}
	\subfigure{
		\begin{minipage}[t]{1\linewidth}
			\centering
			\includegraphics[width=2.7in]{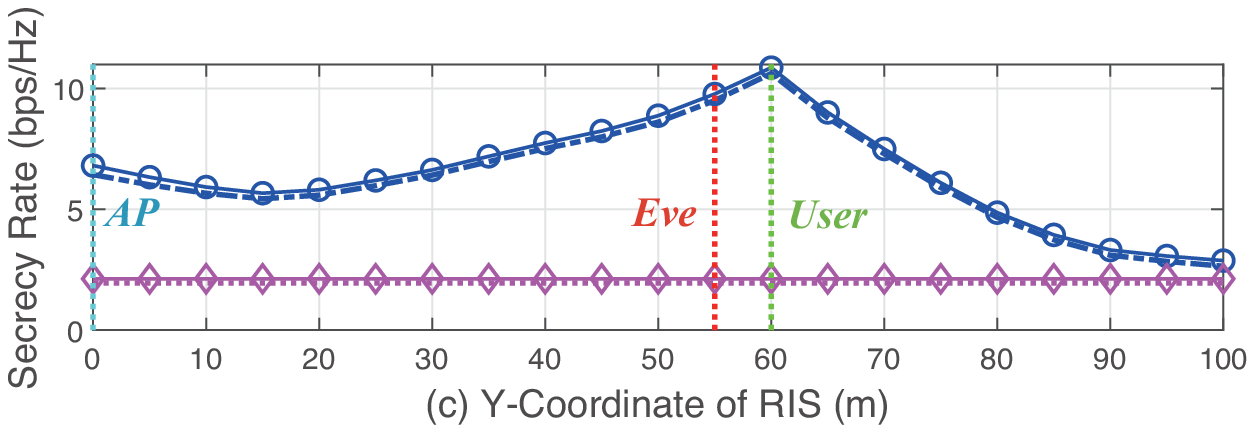}
		\end{minipage}
	}
	\centering
	\caption{ESR and LESR versus the $Y$-Coordinates of Eve, user and RIS when $P_{\rm max} = 5$dBm and $N=160$.}
	\label{fig4}
\end{figure}

\begin{figure}[t]
	\centering
	\subfigure[Same setting as TABLE \ref{tab1}.]{
		\begin{minipage}[t]{0.465\linewidth}
			\centering
			\includegraphics[width=1.5in]{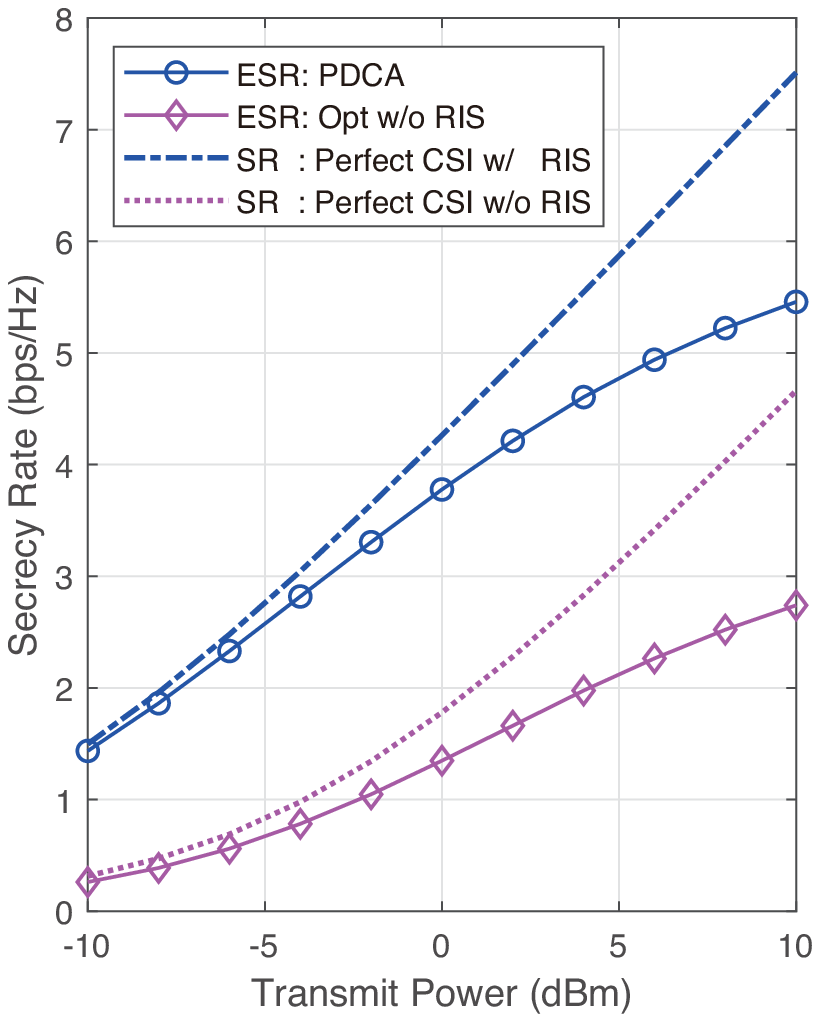}
		\end{minipage}
	}
	\subfigure[Only NLoS path between RIS and user/Eve.]{
		\begin{minipage}[t]{0.465\linewidth}
			\centering
			\includegraphics[width=1.5in]{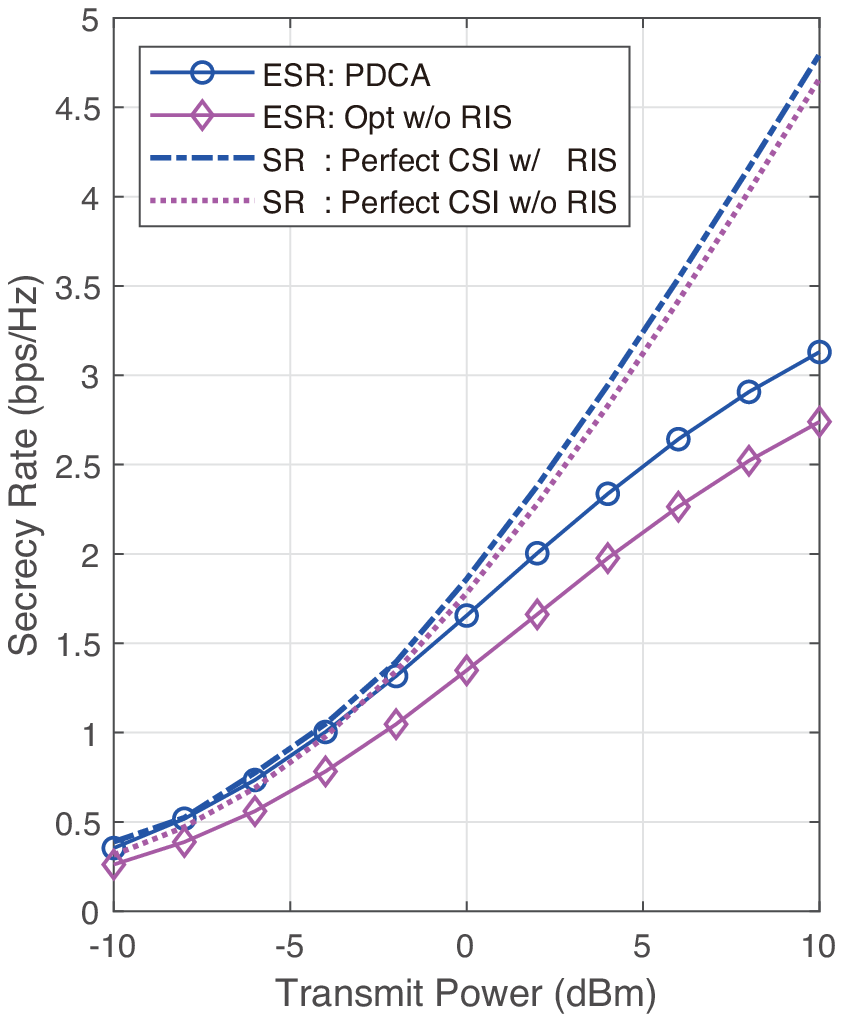}
		\end{minipage}
	}
	\centering
	\caption{ESR and SR versus $P_{\rm max}$ when $N=32$}
	\label{fig5}
\end{figure}

As depicted in Fig. \ref{fig2}, we consider an uniform linear array (ULA) at the AP and an uniform planar array (UPA) equipped with $N=N_y N_z$ reflecting elements at the RIS. For ease of exposition, $N_y=16$ is fixed and $N_z$ is increased linearly according to $N$. The $z$-coordinate of AP, RIS, user and Eve are set to $0$. We assume the AP-user/Eve direct channel follows Rayleigh fading while the RIS-assisted channels follow Rician fading as in\cite{HuayanGuo20}. The simulation parameters are summarized in Table \ref{tab1}. In particular, the path loss exponent and Rician $K$-factor of AP/RIS-user/Eve channel are set according to the 3GPP Urban Micro-cell(UMi) propagation environment\cite{3GPP} and those of AP-RIS channel are the same as those in\cite{QingqingWuTWC}.

We consider following schemes for comparison: \textbf{1) PDCA:} the proposed PDCA joint-design scheme; \textbf{2) AO:} the alternating optimization (AO) scheme proposed in\cite{MiaoCui19}. To be more specific, on the one hand, the RIS phase shifts design problem is solved by semidefinite relaxation (SDR) with Gauss randomization technique. On the other hand, the transmit beamforming design problem is derived in closed-form by leveraging the solution of the generalized Rayleigh quotient problem\cite{AshishKhisti10}; \textbf{3) Opt w/o RIS:} the beamformer is designed to maximize the LESR in the absence of RIS. By substituting the solution of the LESR maximization problem into formula (7), ESR is computed by Monte Carlo simulation through taking average of $10^5$ i.i.d. random $\mathcal{H}_{\rm E} = \{\B h_{\rm AE}, \B h_{\rm IE}\}$ realizations.

In Fig. \ref{fig3}, we compare the ESR and LESR of all schemes versus $P_{\rm max}$ and $N$, respectively. It can be seen that the proposed PDCA scheme achieves the highest ESR among all schemes. In Fig. \ref{fig3}(a), more than $1$bps/Hz performance gain is achieved by adopting PDCA compared with AO. It is observed from Fig. \ref{fig3}(b) that the performance gap between ESR and LESR of same scheme is quite small (less than $0.3$bps/Hz), which demonstrates that the LESR serves as a relatively tight lower bound. The ESR of both RIS-assisted schemes are increasing with $N$, owing to the \emph{beamforming gain} and \emph{aperture gain}\cite{QingqingWuTWC} provided by RIS.

In Fig. \ref{fig4}, we compare the ESR and LESR versus the $Y$-coordinates of Eve, user and RIS, respectively. In the optimal beamforming scheme without RIS, it is natural to see that the ESR increases with the coordinate of Eve and decreases with the coordinate of user. For the RIS-assisted PDCA scheme, Fig. \ref{fig4} (a) shows that the ESR increases with the coordinate of Eve at first and then goes down to a local minimum at 70m (a location near both RIS and user). On the contrary, Fig. \ref{fig4} (b) and Fig. \ref{fig4} (c) demonstrate that the ESR decreases at first and then goes up to a local maximum at 50m (RIS's location) and 60m (user's location), respectively. It inspires us that it is a wise choice to place the RIS near the user out of the consideration for the security performance.

In Fig. \ref{fig5}, two perfect CSI schemes with and without RIS-assisted are simulated to investigate their relationship with corresponding statistical CSI schemes. The RIS-assisted scheme with perfect CSI is also solved by the proposed PDCA method. It can be seen that the performance with perfect CSI serve as upper bounds for the corresponding statistical CSI cases. In Fig. \ref{fig5}(b), we set $\alpha_{\rm IU} = \alpha_{\rm IE} = 3.67$ and $K_{\rm IU} = K_{\rm IE} = 0$ the same as those of the AP-user/Eve channel, which means there is only NLoS path between RIS and user/Eve. Due to the blockage of the LoS path between RIS and user/Eve, it is clear from the Fig. \ref{fig5}(b) that there is nearly no performance gain for the RIS-assisted perfect CSI scheme. Nevertheless, there still exists a prominent ESR improvement between the proposed PDCA scheme and the optimal beamforming without RIS scheme, owing to the exploitation of the Eve-related statistical CSI.

From the simulation results above, it is satisfying to see that the security performance of the RIS-assisted PDCA scheme is better than the optimal beamforming scheme without RIS, which proves the power of RIS in physical layer security.

\section{Conclusion}
We studied the RIS-assisted secure transmission with the knowledge of statistical CSI of eavesdropper. A lower bound of ESR is derived. Therefore, the stochastic ESR maximization problem is converted into a deterministic LESR maximization problem. By leveraging the proposed PDCA algorithm, the solution of this non-convex optimization problem is guaranteed to converge to a KKT point. Simulation results demonstrate that the security performance achieved by the proposed PDCA scheme is better than the commonly adopted AO scheme. The performance under perfect CSI settings are also given for comparison to acquire more useful insights.

\appendices
\section{Proof of Proposition 1}
By the definition of ESR in (7), we have
\begin{align}
	\bar{R}_{\rm S} &\geq \left[ R_{\rm U} - \mathbb{E}_{\mathcal{H}_{\rm E}} \left[ R_{\rm E} \right] \right]^+\notag\\
	&\geq \left[ R_{\rm U} - \log_2 \left(1 + \frac{1}{\sigma_{\rm E}^2} \mathbb{E}_{\mathcal{H}_{\rm E}} \left[\left|\left(\B{h}_{\rm IE}\CT \B\Phi \B{H}_{\rm AI} + \B{h}_{\rm AE}\CT \right) \B{w}\right|^2\right]\right) \right]^+\notag\\
	& = \bar{R}_{\rm S}^{\rm lb}
\end{align}
where the first inequality is due to the convexity of $\left[\cdot\right]^+$, and the second inequality is due to Jensen's inequality. The equality holds because of the independence between $\B h_{\rm AE}$ and $\B h_{\rm IE}$ together with $\mathbb{E} [|\tilde{h}_{{\rm AE},i}|^2] = \mathbb{E} [|\tilde{h}_{{\rm IE},j}|^2] = 1, \ i=1,\ldots,M, \ j=1,\ldots,N$. 

\ifCLASSOPTIONcaptionsoff
\newpage
\fi

\bibliographystyle{IEEEtran} 

\bibliography{Ref}

\end{document}